\documentclass{article}

\usepackage{amsmath}
\usepackage{amsfonts}
\usepackage{amssymb}
\usepackage{enumitem}
\usepackage{float} 

\usepackage{arxiv}

\usepackage[utf8]{inputenc} 
\usepackage[T1]{fontenc}    
\usepackage{hyperref}       
\usepackage{url}            
\usepackage{booktabs}       
\usepackage{amsfonts}       
\usepackage{nicefrac}       
\usepackage{microtype}      
\usepackage{lipsum}
\usepackage{fancyhdr}       
\usepackage{graphicx}       
\graphicspath{{media/}}     

\title{Pool samples to efficiently estimate pathogen prevalence dynamics}

\author{
  Braden Scherting \\
  Dept. Mathematical Sciences \\
  Montana State University \\
  MT, USA \\
  \texttt{bradenscherting@montana.edu} \\
   \And
  Alison Peel \\
  Centre for Planetary Health and Food Security \\
  Griffith University \\
  Queensland, AU\\
  \And
  Raina Plowright \\
  Dept. Microbiology and Immunology \\
  Montana State University \\
  MT, USA\\
  \And
  Andrew Hoegh \\
  Dept. Mathematical Sciences \\
  Montana State University \\
  MT, USA \\
}

\begin{document}
\maketitle

\begin{abstract}
Estimating the prevalence of a disease is necessary for evaluating and mitigating risks of its transmission within or between populations. Estimates that consider how prevalence changes with time provide more information about these risks but are difficult to obtain due to the necessary sampling intensity and commensurate testing costs. We propose pooling and jointly testing multiple samples to reduce testing costs and use a novel nonparametric, hierarchical Bayesian model to infer population prevalence from the pooled test results. This approach is shown to reduce uncertainty compared to individual testing at the same budget and to produce similar estimates compared to individual testing at a much higher budget through two synthetic studies and two case studies of natural infection data.
\end{abstract}

\keywords{Bayesian methods, environmetrics, nonparametric methods, sampling, time series}

\section{Introduction}

The risk of pathogen spillover from animals to people (zoonotic spillover) is a function of the prevalence of the pathogen in animal populations in space and time \cite{becker2021ecological}. Therefore, understanding patterns of pathogen prevalence is fundamental to mitigating zoonotic disease events but the intensive testing required to do so is often prohibitively costly. Testing costs can be reduced by testing pools of samples, but it is difficult to estimate prevalence from pooled samples over time. In this paper, we develop inferential tools for efficiently estimating pathogen prevalence time series processes from pooled data. These methods can be applied to estimate and mitigate the risk of zoonotic spillover and therefore provide a tool for pandemic prevention.

Prevalence estimation is central to disease surveillance in reservoir host populations \cite{kuiken2005pathogen}, however, because true prevalence is not observed except in trivial settings, estimating prevalence requires sampling, testing, and inference. Modeling prevalence dynamics can lead to more precise predictions of prevalence and, thereby, spillover but most approaches require intensive sampling and testing efforts \cite{plowright2019sampling}. Furthermore, the diversity and complexity of additional factors that contribute to spillover risk limit the availability of resources dedicated to sampling and testing for prevalence estimation \cite{plowright2017pathways}. Efficient allocation of testing resources coupled with dynamic modeling will enable better identification of current and future pathogenic hazards. 

A popular strategy for reducing testing costs when prevalence is expected to be low is to test multiple samples jointly through a procedure known as group testing or pooled testing. Pooled testing proceeds by randomly constructing groups or pools of individual samples and expending a single test on each pool. With a perfect test, a positive result implies one or more individuals in the pool is infected, and a negative result implies all individuals are not infected. These implications are complicated slightly by imperfect tests and pooling-induced dilution, though these limitations are usually outweighed by benefits in practice. Individual diagnoses could then be obtained by retesting members of pools that test positive. Originally developed to scale up the US military's ability to screen recruits for syphilis, the strategy has seen a resurgence in popularity due to the COVID-19 pandemic \cite{garg2021evaluation}. When applied to syphilis screening and standard COVID-19 testing, group testing is used in a diagnostic capacity. However, when individual results are not required and pathogen prevalence is the quantity of interest, as is the case in surveillance of reservoir host populations, follow-up testing can be omitted, reducing testing costs further. \cite{hoegh2021estimating} found that pooled testing can be used to estimate prevalence as accurately and more efficiently than individual testing. However, their results are not directly applicable to observations that exhibit natural temporal correlation, such as multiple samples collected from the same site at different times. 

Knowledge of the distribution of infection with respect to time enables better understanding of the factors that lead to high prevalence and therefore prediction of spillover events. However, testing limitations often beget inferences with limited temporal scope or resolution and wide uncertainty intervals. Efficient sampling and inference strategies for estimating prevalence therefore must be extended to account for and leverage correlation among temporally-indexed observations. Such an extension should achieve interpolation between observation times and smoothing of noisy observations. Interpolation generalizes pointwise inferences to unobserved, intervening times and enables improved qualitative and quantitative characterizations of trends that may not be apparent from pointwise inferences alone. Independent estimates of prevalence at different points in time are noisy; under the assumption that nearby events (in space or time) are more closely related than distant ones, smoothing de-noises estimates by sharing information between neighboring observations. Smoothing can also produce more comprehensive representations of uncertainty. In the absence of direct observation, prevalence is modeled as a latent process with a transformed Gaussian process (GP) prior and is related to pooled test results through a Bayesian hierarchical model. 

Our goals are twofold: 1) establish the ability of pooled data to recover true, underlying time-varying prevalence and 2) demonstrate that pooled testing can reduce testing costs without degrading estimates or, alternatively, generate more precise inferences at a fixed testing budget. The remainder of the article is organized as follows. In Section 2, we introduce the motivating data and the hypothetical pooling design. In Section 3, we briefly review pooled testing and GP methodology, introduce the modeling framework, and comment on relevant computational considerations. Section 4 reports results from synthetic studies demonstrating recovery of true prevalence and case studies evidencing the relative efficiency of pooled data. The results from the motivating case studies are reported in Section 5, along with additional consideration of sampling variability inherent to our studies. In Section 6 we discuss implications of the present work and future directions.  

\section{Data}

We consider two motivating case studies. The first is an example of pathogen surveillance among a reservoir host population where low-cost prevalence estimation is necessary for determining trends and designing interventions. The second considers disease surveillance in a human population where knowledge of prevalence may instead be used to inform institutional and public health policies.  
\subsection{Case study 1: Congo Basin bats}
\cite{kumakamba2021coronavirus} report the results from 3,561 coronavirus surveillance tests performed on samples from bats, rodents and primates over 12 years collected throughout the Republic of Congo and Democratic Republic of Congo. We restrict our attention to samples collected from bats. Sampling efforts occurred roughly monthly, and each site was visited roughly twice per year. An approximately two-year lapse in sampling occurred between 2013 and 2015, so, absent the ability to meaningfully interpolate over two years, we evaluate our methods on a subset of the total sampling interval comprising 752 individual tests performed between 24 August 2015 and 20 July 2018. If an individual was tested twice (e.g., using both fecal and saliva samples), we record it only once and deem an individual positive if any of the multiple samples tests positive. Though the date that each sample was collected is recorded, sampling efforts span multiple days. Because we do not expect prevalence to change meaningfully over the short window of days within which sampling occurs, we aggregate results from the same location that occur within $<10$ days of each other and record the date as the date of the first sample collection date in the sampling effort. This procedure is consistent with the description of sampling provided by the authors and reflects the reality of sampling reservoir hosts. All processing described here was performed in advance of any analysis and is identical between pooled and individual testing analyses.

\subsection{Case study 2: asymptomatic testing at Notre Dame}

In the face of the COVID-19 pandemic, many higher education institutions in the United States implemented on-campus diagnostic and surveillance testing during the 2020/21 academic year. This, in combination with other public health precautions, was intended to limit transmission of SARS-CoV-2 through early detection and subsequent isolation \cite{fox2021response}. The University of Notre Dame, a medium-sized university in Indiana, was one such institution. We examine results from nasal swab and saliva tests administered to asymptomatic students on campus on a daily basis between August 3 and December 16, 2020, published publicly in dashboard format on the university's website \cite{ndcovid}. The 81,872 tests identified 248 positive cases, for an overall positive rate of 0.30\%. In this scenario, individual diagnostic results were sought after, as is often the case when testing human populations. However, prevalence estimation remains relevant. In humans, knowledge of population prevalence is used to inform decision-making (e.g., whether to revert to online instruction). Pooled test results from initial testing may also be used to direct later, follow-up testing to obtain individual diagnoses. For example, testing bottlenecks may be combated by prioritizing pooled testing and prevalence estimation for decision-making, and follow-up testing of positive pools can occur when testing resources become available. 

\subsection{Hypothetical pooling design}

Both data sets contain individual test results. To evaluate methods for estimating true prevalence from tests performed on pooled samples, we construct hypothetical pools by randomly grouping individual results and computing pooled test results. A pool is positive if one or more individuals tested positive in the original data and negative otherwise, thereby censoring individual-level data.

Broadly, testing costs can be broken down into a fixed cost that is invariant to the number of tests and a variable cost that scales with the number of tests performed. Reductions in variable testing costs can be achieved through two parallel approaches: 1) reducing the number of tests needed to obtain similarly precise and accurate estimates (obtaining the same results at lower cost), or 2) improving the precision and accuracy of estimates at a fixed budget (obtaining better results at the same cost). The first approach is evaluated by comparing prevalence estimates obtained from pooling to prevalence estimates obtained from the original data. The second approach is evaluated by comparing pooled estimates to estimates obtained from a subsample of the original data representing the same number of tests (i.e., fixed variable cost). Let $m$ be the number of individuals in a given pool. Prevalence estimates that use all available data are labeled $m=1$ and are regarded as the best approximation to true prevalence and therefore the best known alternative to the proposed method. Estimates from subsampled, individual tests are labeled $m=1^*$, and estimates from pool tests are labeled with their respective pool sizes (e.g., $m=5$). Synthetic pooling and subsampling is performed within sampling instances or time steps. For example, if 17 individuals were originally sampled at time $t_i$, then the $m=1$ estimate is based on all 17 individuals, the $m=3$ estimate is based on 5 pools of size 3 and 1 of size 2, and the $m=1^*$ estimate is based on 6 subsampled individuals, because 6 pools and therefore 6 tests are used in the pooling setup. When multiple pool sizes are considered simultaneously, the larger testing budget is used to construct the $m=1^*$ subsample. In the Congo Basin bat study, pool sizes 3 and 5 are used. In the Notre Dame study, pool sizes 5 and 10 are used. 

\section{Model Formulation}

The methods described here can be used to infer the population prevalence or probability that an individual selected randomly from a specified population is infected with an infectious disease of interest, which is denoted $p$, at any given time within a specified interval. We observe only whether or not a sample tests positive for the pathogen and the sample collection date. Under appropriate conditions, testing pools or groups comprising samples from multiple individuals confers considerable reductions in testing costs. Therefore, a hierarchical model is used to relate the binary or count data obtained through pooled testing to the probability that a pool tests positive, $\pi$, and to relate this pool probability back to individual probability, $p$. 

\subsection{Population prevalence and pool probability}

In possession of test results from pooled samples, estimating prevalence at the pool level is straightforward—one could, for example, employ a beta prior and binomial likelihood and proceed with a conjugate analysis analogous to the procedure described in Section 2.2.1 of \cite{hoegh2021estimating}. However, when prevalence at the individual level is the quantity of interest, any inference on $\pi$ must be related to $p$. To do so, we use the inverse prevalence transformation of \cite{colon2001estimating} :
\begin{align}
    \pi = 1 - (1 - p)^m,
\end{align}
where $m$ is, again, the pool size. Inspection reveals that pool probability is high when either prevalence is high or pool size is large. At lower prevalence, $p$ is approximately linear in $\pi$. For high prevalence, however, $p$ is highly non-linear in $\pi$ at which point many values of $p$ imply similar, large values of $\pi$, hindering identifiability. The non-linearity also grows with $m$ (see \cite{colon2001estimating}, Figure 1 ). High prevalence requires small pools; low prevalence permits large pools. \cite{swallow1985group} describes experimental design considerations for optimal pool size; \cite{reilly1996optimal} additionally considers optimality under two-stage designs, where individuals from positive pools are retested. 

\subsection{Latent Gaussian process}

GPs are popular tools for nonparametric regression as they make no assumptions about the functional forms of relationships and few assumptions generally. They can therefore be used as priors over functions to model unknown, possibly non-linear functions automatically. By definition, a GP is a stochastic process $\{X_t\}_{t\in\mathcal{T}}$ where any select variables $X_t=(X_{t_1},\dots ,X_{t_n})^{\intercal}$ are distributed as multivariate Gaussian. The support of a GP is therefore in the real numbers, limiting applications. However, transformations can be used to map this support to different intervals, greatly broadening its uses. The Gaussian distribution function (inverse probit) $\phi(\cdot)$ is one such transformation, commonly used to map GP random variables to $[0,1]$ for use in classification. We adopt a similar approach. 

Because $p_{t_i}\in [0,1]$, it is convenient to model a latent prevalence process $W:=\{W_{t}\}_{t\in\mathcal{T}}$ with real support and transform to obtain $p_t = \phi(w_{t})$. Prevalence, and therefore the form of the latent process, are unknown. Together with assumptions of smoothness, continuity, and real support, this solicits a GP prior on $W$. The only additional user-specified components of a GP are the covariance function and priors over covariance hyperparameters. The form of the covariance function determines the nature of the relationship between process values as a function of time, and proximity in time determines the strength of the covariance between any two process values. In general, we expect process values at similar times to be more similar than process values at distant times. This agrees with our notion of smoothing. Covariance can be computed between any locations to obtain a function value, achieving interpolation. A common, general-purpose covariance function, and the function we use throughout this paper, is given by 
\begin{align} \label{eq_covariance}
Cov(t,t' \mid \theta) =\sigma^2 \exp\left(\frac{-(t-t')^2}{2\ell^2}\right),
\end{align}
known as the exponentiated quadratic covariance function. The hyperparameters $\theta = \{\sigma^2, \ell\}$ control the oscillation speed and amplitude of a sampled process, respectively. Many other covariance functions can be used instead to match the application. The covariance matrix $C$ of our multivariate Gaussian distribution is obtained by letting $C_{ij} =C_{ji} = Cov(t_i,t_j \mid \theta)$, and we write $W_t \sim \mathcal{GP}(\mathbf{0}, C)$. 

Here, we have specified a zero-mean GP; a useful property is that if $F\sim \mathcal{GP}(\boldsymbol{\mu}, C)$ and $G\sim \mathcal{GP}(\boldsymbol{0}, C)$, then $F = G + \boldsymbol{\mu}$. The mean can be modeled independently either as a scalar or some function of covariates and the covariance matrix will capture any residual structure not modeled in the mean. 

\subsection{Sampling models}

Models of test results from pooled samples depend primarily on the size of the pools, because the probability that a pooled sample tests positive is a function of both population prevalence and pool size. Different models arise depending on how pool sizes are assigned. At one extreme, pool sizes and number of pools are determined in advance of sampling individuals and all pools are the same size. At the other extreme, pools of many different sizes are tested, possibly in advance of any analysis. Here, we consider three scenarios representing 1) the general case, where all pools may be different sizes, 2) the ideal case in which all pools are the same size, and 3) an efficient alternative to the general case that represents many realistic situations. Idealized and efficient general models can be regarded as special cases of the general model. In all scenarios, the pool sizes $m$, number of pools $k$, and number of individual samples $n$ are assumed to be known for all pools. Unknown pool sizes could, in principle, be estimated from data, but strong prior information or other forms of regularization would likely be required for identification. In all scenarios, small notational simplifications are obtained when $n$ (and therefore $k$ and $m$, also) is constant in time.

\subsubsection{General}

In the general setting, each pool is given its own probability—no assumptions are needed about the size or number of pools. We use $j\in \{1,\dots,k_{t_i}\}$ to index pools at time $t_i$. The sampling model is
\begin{align}
    \begin{split}
        Y_{t_i, j} &\sim \text{Bernoulli}(\pi_{t_i,j}) \\
        \pi_{t_i,j} &= 1- (1-p_{t_i})^{m_{t_i,j}},
    \end{split}
\end{align}
where $Y_{t_i,\cdot}$ is a vector of binary variables indicating which pools of samples test positive at time $t_i$. At each time, $\pi$ is indexed by pools, but $p$ is constant. We use this model in the Congo Basin bat surveillance example. If $m_{t_i,a} = m_{t_i,b}$ for all $a,b\in \{1,\dots,k_{t_i}\}$, this model simplifies to the idealized model described next. 

\subsubsection{Ideal}

In the ideal setting, all pools at time $t_i$ contain $m_{t_i}$ individual samples, and $k_{t_i}$ pools are tested. This requires $m_{t_i}\times k_{t_i} = n_{t_i}$. In the absence of pool-level covariates, all pools at a given time have the same probability of testing positive because both $m_{t_i}$ and $p_{t_i}$ are the same for all pools. The sampling model is 
\begin{align}
    \begin{split}
        Y_{t_i}&\sim \text{Binomial}(k_{t_i}, \pi_{t_i})\\
        \pi_{t_i} &= 1- (1-p_{t_i})^{m_{t_i}},
    \end{split}
\end{align}
where $Y_{t_i}\in\{0,1,\dots,k_{t_i}\}$ represents the number of pools that test positive at $t_i$. Synthetic examples follow this framework. This model is not appropriate when pools are different sizes, as is the case when the pool size does not divide the number of individual samples. 

\subsubsection{Efficient general}

The total number of individual samples may not be perfectly divisible by the chosen pool size. For example, if as many samples as possible are collected rather than a predetermined number, it is unlikely that $m$ divides $n$. Suppose we test pools no larger than $m^*$. We can use the idealized model when $m^*$ divides $n_{t_i}$, and the general model otherwise. However, when $k$ is on the order of tens or hundreds it is computationally advantageous to model pool structure more deliberately. When $m^*$ does not divide $n_{t_i}$, test $k_{t_i}-1$ pools of size $m^*$ and one pool of size $(n_{t_i} \;\mathrm{mod}\; m^*) = m_{t_i}$. This motivates the final sampling model:
\begin{align}
    \begin{split}
        Y_{t_i,1}&\sim \text{Binomial}(k_{t_i}-1, \pi_{t_i, 1})\\
        &\pi_{t_i, 1} = 1- (1-p_{t_i})^{m^*};\\
    \end{split}
\end{align}
\begin{align}
    \begin{split}
        Y_{t_i,2}&\sim \text{Bernoulli}(\pi_{t_i, 2})\\
        &\pi_{t_i, 2} = 1- (1-p_{t_i})^{m_{t_i}}.
    \end{split}
\end{align}
Here, the response $Y_t=(Y_{t,1}, Y_{t,2})^{\intercal}$ is a multivariate random vector comprising the number of positive pools among those of size $m^*$ and a binary variable indicating the test result of the final, odd-sized pool. Because the number of individual samples may vary over time in this scenario, it may be the case that $n_{t_i} < m^*$ or $k_{t_i} \times  m^* = n_{t_i}$. This model handles both situations naturally. In the former case, we have $Y_{t_i,1}\sim \text{Binomial}(0, \pi_{t_i, 1})$ and in the latter, $Y_{t_i,2}\sim \text{Bernoulli}(0)$, which imply $Pr(Y_{t_i,1} = 0) = Pr(Y_{t_i,2} = 0) = 1$; in effect, either the ideal or general model is used automatically. Obviously, $y_{t_i,1}$ and $y_{t_i,2}$ must be coded as zeros in the respective situations. This formulation can reduce the number of likelihood terms by orders of magnitude, lending a substantial speed-up. We use this model in the Notre Dame COVID-19 testing example, in which more than 80,000 data are analyzed. 

\subsection{Hierarchical specification}

Let $t = \{t_1, t_2, \dots, t_n\}$ be an index set of observation times in the observation interval $\mathcal{T}$ with observed responses $y_t = \{y_{t_1}, \dots, y_{t_n}\}$ where $y_{t_i}\in \{1,\dots,k_{t_i}\}$, corresponding to the ideal scenario described above. Additionally, let $W_t$ be the set of latent prevalence values at times $t$. The hierarchical model is
\begin{align}
\begin{split} \label{hier}
    y_{t_i} &\sim \text{Binomial}(k_{t_i}, \pi_{t_i}) \\
    \pi_{t_i} &= 1 - (1 - p_{t_i})^{m_{t_i}} \\
    p_{t_i} &= \phi(W_{t_i} + \mu) \\
    W_t &\sim \mathcal{GP}(\mathbf{0}, C) \\
    \mu &\sim p(\mu) \\
    \theta &\sim p(\theta).
\end{split}
\end{align}
Replace the first two lines with the appropriate sampling model in non-ideal scenarios. In the absence of covariates, we model a scalar mean $\mu$. Instead modeling the mean as a linear combination of predictors $\boldsymbol{\mu}= X\beta$ would be a natural extension. 

The model used to infer prevalence from individual samples is the same as \eqref{hier} with the prevalence transformation omitted and the sampling model
\begin{align*}
    y_{t_i} \sim \text{Binomial}(k_{t_i}, p_{t_i}),
\end{align*}
where $k_{t_i}$ is now the number of individuals tested at $t_i$ and $y_{t_i}\in \{1,\dots,k_{t_i}\}$ is the observed number of infected individuals. 

\subsection{Priors and computation}

In the interest of propagating uncertainty as thoroughly as possible, we perform fully-Bayesian inference on all model parameters and hyperparameters by use of Markov chain Monte Carlo (MCMC). Probabilistic inference on GP hyperparameters is challenging in this setting due to the posterior dependence between latent variables and hyperparameters; the coupling can be resolved by intensive sampling, but the $\mathcal{O}(N^3)$ scaling behavior of GP prediction renders such intensive, repeated computation impractical. The data sets considered here are sufficiently small that inference in Stan \cite{carpenter2017stan} remains practical. For larger data sets, the dependence structure must be addressed directly. Several methods exist; see \cite{filippone2013comparative} for a review.

We use normal and half-normal priors on $\mu$ and $\sigma$, respectively. Values of $\ell$ less than the shortest time between observations (\emph{low}) or greater than the total observation interval (\emph{high}) are not identified. Accordingly, an inverse-gamma prior on $\ell$ weakly inform the range of plausible values between \emph{low} and \emph{high}. 

\section{Synthetic studies}

Dynamic prevalence estimation from pooled samples is first evaluated through two studies of simulated data. The simulated studies are used to evaluate the proposed method's ability to recover known, underlying prevalence in settings similar to those of the case studies. The simulated studies are designed to reflect characteristics present in the data used in the applications. Clearly, it is not possible to simulate a perfect match to the unobserved prevalence processes that gave rise to the observed data, but we approximately match the order of magnitude, overall trend, and sampling design. In the first study, true prevalence ranges from $0$ to $0.12$ and oscillates slowly in that range. The prevalence is 'observed' at 25 evenly spaced times within a 1000 day interval. 45 individual test results are simulated for each observed prevalence and 15 pools of size 3 ($k=15$, $m=3$) are constructed. In the second study, true prevalence does not exceed 0.05 and tests are simulated every day for 150 days. 500 individual test results are simulated each day and pool size $m=10$ is used, so 50 pools are required each day ($k=50$, $m=10$). Here, a larger pool size is tolerated because prevalence is low. This would not be known\emph{ a priori}, but may be available through domain knowledge, past analyses, or initial testing. In both studies, $m=1$ is informed by $k\times m$ individual data, $m=1^*$ is informed by $k$ individual data, and $m>1$ are informed by $k$ pools of size $m$. 

\begin{figure}[ht]
\centerline{\includegraphics[width=500pt,height=283pt]{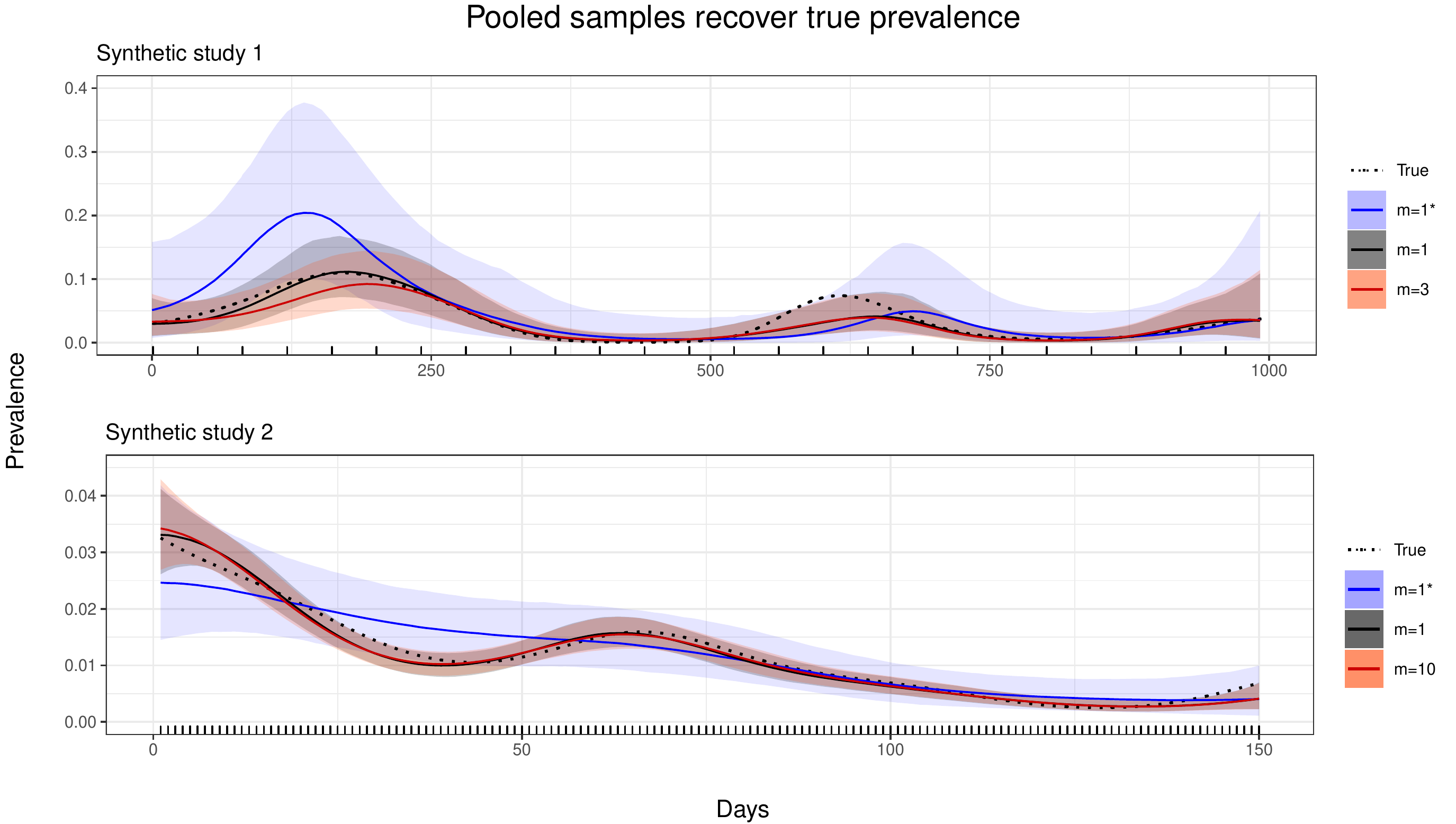}}
\caption{Estimated median curves and 95\% credible regions are displayed for each synthetic study under each sampling strategy. Rug marks indicate when sampling occurred. In both cases, the pooled estimates closely track true, unobserved prevalence (black, dotted) and $m=1$ estimates (black, solid). Estimates from limited individual sampling (blue, solid) are less precise and fail to accurately recover true prevalence. \label{fig1}}
\end{figure}

\begin{table}[ht]
\caption{GP hyperparameter estimates for synthetic studies ($\mathbf{1}$ and $\mathbf{2}$). Estimated posterior means and 95\% CIs are reported for each parameter and sampling regime.} \centering
\begin{tabular}{ r | c | c c c } \toprule
$\mathbf{1}$ & True & $m=1$ &   $m=3$ &  $m=1^*$\\
\midrule
$\ell$ & 100 & 89.7 (53.8, 136.9)  & 91.3 (55.6, 139.1) & 92.8 (50.0, 164.6)  \\
$\sigma$ & 0.5 & 0.59 (0.31, 1.00) & 0.56 (0.29, 0.98) & 0.67 (0.32, 1.18) \\
$\mu$ & -2 & -1.99 (-2.46,  -1.50) & -2.00 (-2.46, -1.53) & -1.90 (-2.50, -1.38) \\
\hline 
\hline
$\mathbf{2}$ & True & $m=1$ &   $m=10$ &  $m=1^*$\\
\midrule
$\ell$ & 25 & 23.3 (16.8, 32.4)  & 24.4 (17.6, 34.4) & 34.8 (18.9 64.7)  \\
$\sigma$ & 0.25 & 0.38 (0.20,  0.71) & 0.39 (0.21, 0.72) & 0.36 (0.15,  0.74) \\
$\mu$ & -2.33 & -2.25 (-2.63, -1.80) & -2.24 (-2.63, -1.80) & -2.25 (-2.68, -1.79 ) \\ \bottomrule
\end{tabular}
\label{t12}
\end{table}

In both simulated studies, there is strong agreement between estimated curves for individual ($m=1$) and pooled ($m=(3,5)$) data, displayed in Figure \ref{fig1}. In the top panel, pooled and individual curves track the true prevalence closely with a clear but slight exception around day 625. The budgeted individual curve ($m=1^*$) is both less precise and less accurate in general. The second study (Figure \ref{fig1}, panel 2), in which much lower prevalence is estimated from more data over a shorter interval, provides even more compelling results. True prevalence is estimated well by individual and pooled data, and the two are largely indistinguishable. Furthermore, using budgeted individual data grossly over-smooths the curve, reducing the process to an approximately linear function. Notably, where true prevalence is 0, pooled results are disproportionately informative because it is known that exactly $k\times m$ individuals are negative, whereas individual results at the same testing budget indicate only that $k$ individuals are negative—pooling lends greater precision through sample size when prevalence is near zero. Table \ref{t12} supports these results; compared to $m=1$ and pooled estimates, $m=1^*$ estimates are either less accurate, less precise, or both, with the exception of $\mu$ in the second synthetic study. Together, these simulation studies demonstrate that 1) true, underlying prevalence can be estimated using pooled sample data together with the proposed hierarchical model, 2) for a fixed number of tests, use of pooled sample data confers substantial gains in precision and accuracy, and 3) results from pooled data are consistent with results from individual data, which require $m$ times as many tests and therefore $m$ times higher variable cost. Having established that true prevalence can be efficiently estimated from pooled data, we proceed with analyzing the data described in Section 2.

\section{Results}

\subsection{Case study 1: Congo Basin bats}

\begin{figure}[h!]
\centerline{\includegraphics[width=500pt,height=227pt]{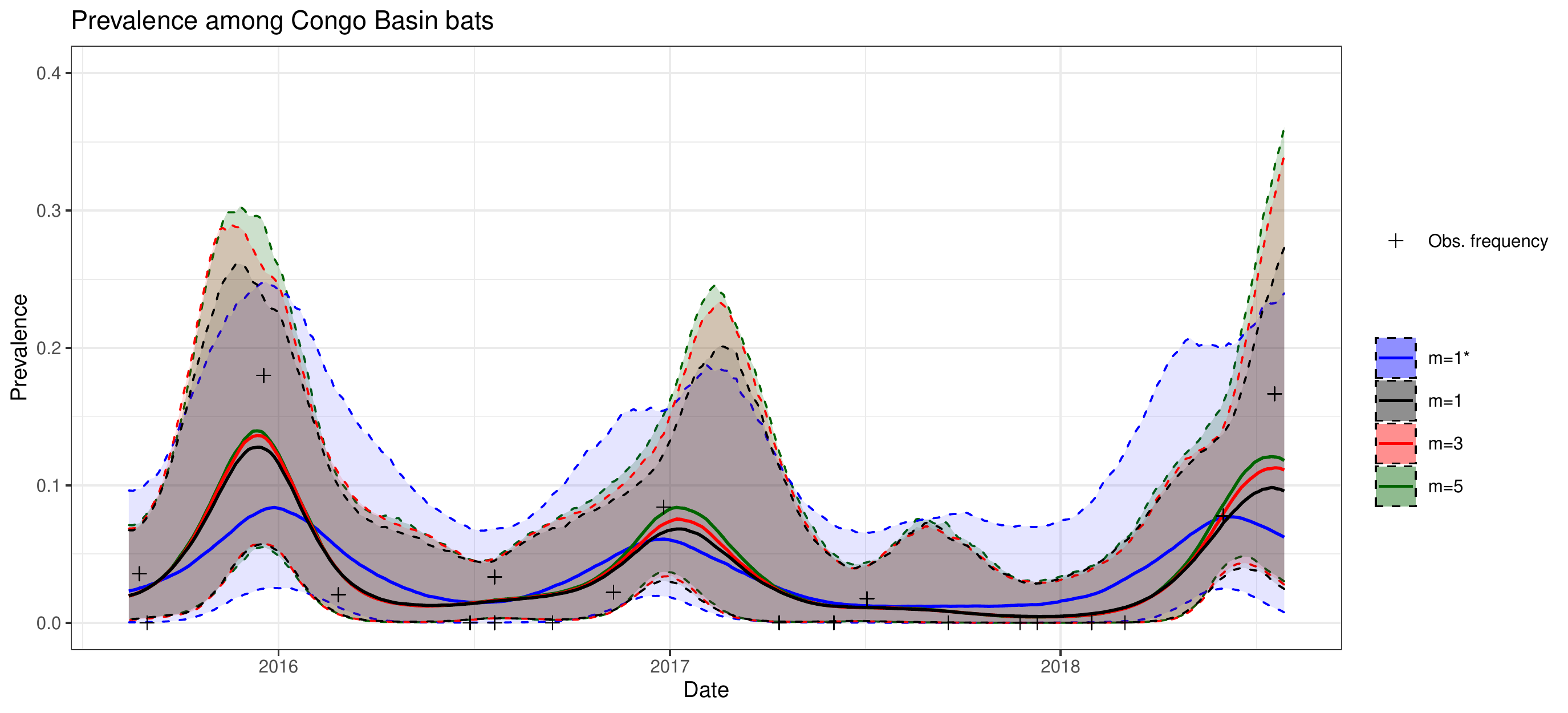}}
\caption{Prevalence among Congo Basin bats considered in this study is estimated in four ways; medians and 95\% credible regions are displayed for each. Universal individual testing estimates (black) and pooled estimates (red, green) exhibit high visual similarity; pooling introduces slight, additional uncertainty. Estimates from limited individual testing (blue) capture the same overall trend but deviate from other estimates and possess more uncertainty throughout. \label{fig2}}
\end{figure}

\begin{table} [h]
\caption{GP hyperparameter estimates under each sampling regime for case study 1. Estimated posterior means and 95\% CIs are reported for each parameter.} \centering
\begin{tabular}{ r | c   c   c   c  } \toprule
        & $m=1$ & $m=3$ & $m=5$ &  $m=1^*$\\
\midrule
$\ell$ & 77.3 (45.3, 132.5) & 75.9 (45.2, 122.7)  & 76.4 (44.5, 127.3) & 89.1 (48.9, 167.8)  \\
$\sigma$ & 0.63 (0.30, 1.08) & 0.65 (0.31, 1.12) & 0.67 (0.34, 1.12)  & 0.56  (0.05, 1.12) \\
$\mu$ & -1.95 (-2.47, -1.48) & -1.94  (-2.47,  -1.48) & -1.93 (-2.46,  -1.43) & -1.93 (-2.56,  -1.43) \\
\bottomrule 
\end{tabular}
\label{t3}
\end{table}

Pooled estimates ($m=3$ and $m=5$), representing 3- and 5-fold reductions in the number of tests, closely match the $m=1$ curve obtained through universal individual testing, both in median and uncertainty intervals (Figure \ref{fig2}). The $m=1^*$ curve, which has equal testing cost to $m=3$ and greater cost than $m=5$, is both less precise and a poorer fit to the best alternative, $m=1$.
Table \ref{t3} tells a similar story. Parameter estimates from pooled data closely match $m=1$ estimates in terms of both mean and 95\% interval endpoints. Estimates from $m=1^*$ are either biased (from $m=1$) or less precise. 

\subsection{Case study 2: asymptomatic testing at Notre Dame}

\begin{figure}[h!]
\centerline{\includegraphics[width=500pt,height=227pt]{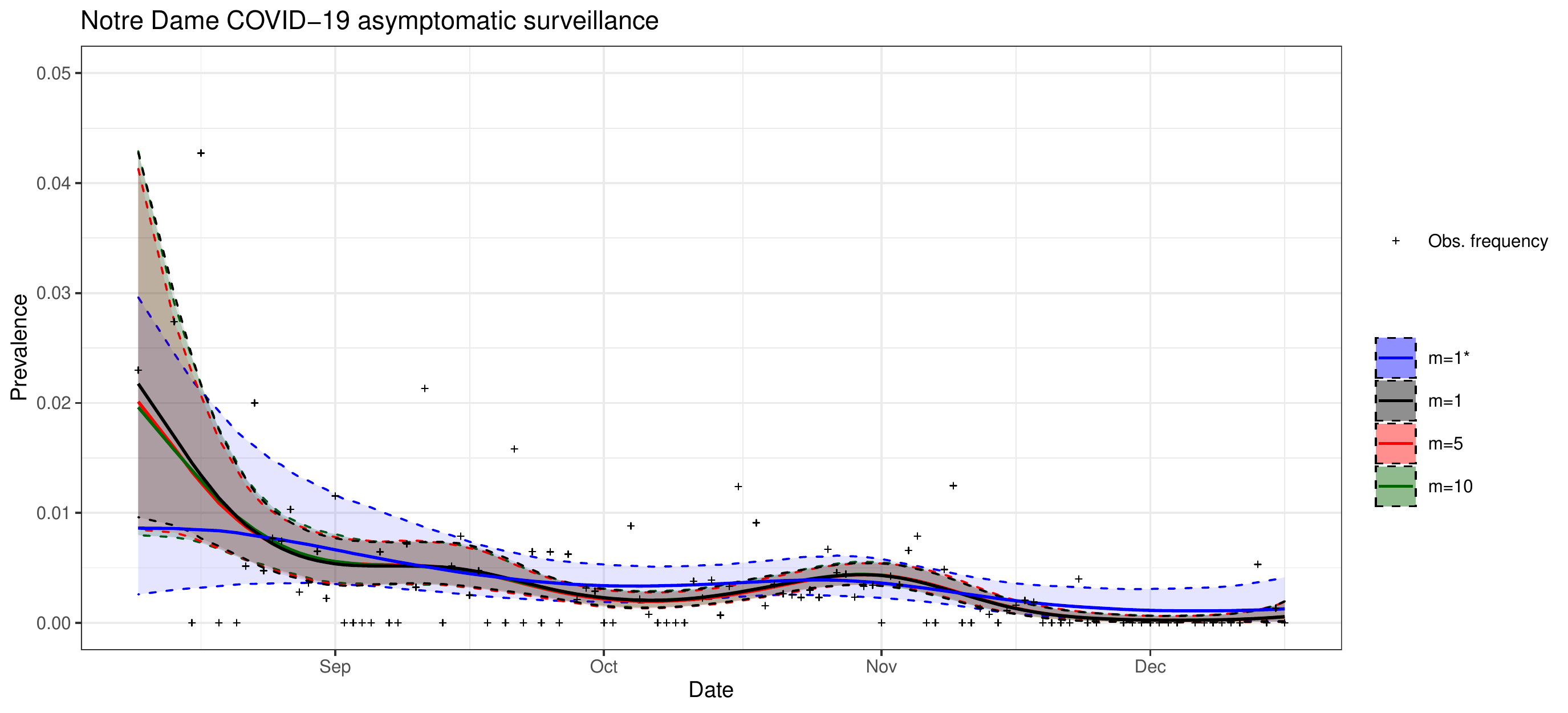}}
\caption{Four estimated median prevalence curves and associated 95\% credible regions of prevalence among asymptomatic individuals at Notre Dame are displayed. For most time points, universal individual (black) and pooled (red, green) estimates are visually indistinguishable. Estimates from limited individual testing (blue) are, again, deviant and more uncertain, on average.\label{fig3}}
\end{figure}

\begin{table}[h]
\caption{GP hyperparameter estimates under each sampling regime for case study 2. Estimated posterior means and 95\% CIs are reported for each parameter.} \centering
\begin{tabular}{ r | c   c   c   c  } \toprule
        & $m=1$ & $m=5$ & $m=10$ &  $m=1^*$\\
\midrule
$\ell$ & 19.7 (13.4, 27.8) & 20.2 (13.5, 28.3)  & 20.3 (13.8, 27.8) & 27.4 (16.3, 47.4)   \\
$\sigma$ & 0.60 (0.32, 1.08) & 0.60 (0.32, 1.05) & 0.58 (0.31, 1.03)  & 0.39 (0.10, 0.88) \\
$\mu$ & -2.49 (-3.02, -1.84) &  -2.49 (-3.00, -1.81)  & -2.50 (-3.01, -1.80)  & -2.57 (-3.01, -2.02) \\
\bottomrule 
\end{tabular}
\label{t4}
\end{table}

Across all testing strategies, the number and frequency of tests lends greater precision, compared to the previous example. Figure \ref{fig3} displays estimated curves from $m=5$ and $m=10$, which are nearly indistinguishable from the universal individual testing curve, $m=1$. Due to the low prevalence and testing intensity, pooled testing is extremely efficient—an order-of-magnitude reduction in testing costs affects estimates almost imperceptibly. However, uncertainty intervals for $m=1^*$ are again far wider and the estimates are clearly over-smoothed. Table \ref{t4} also indicates that pooled estimates recover $m=1$ estimates with high fidelity, but $m=1^*$ estimates differ considerably in expectation and precision. 

Observed frequencies at most times are consistent with the posterior predictive distribution, but very low observed frequencies coupled with large sample sizes are represented only in the extreme left tails of predictive distributions at the respective times. Incorporating demographic information or correlation among individuals tested at the same time may resolve this.

\subsection{Sampling variability}

\begin{figure}[h]
\centerline{\includegraphics[width=500pt,height=296pt]{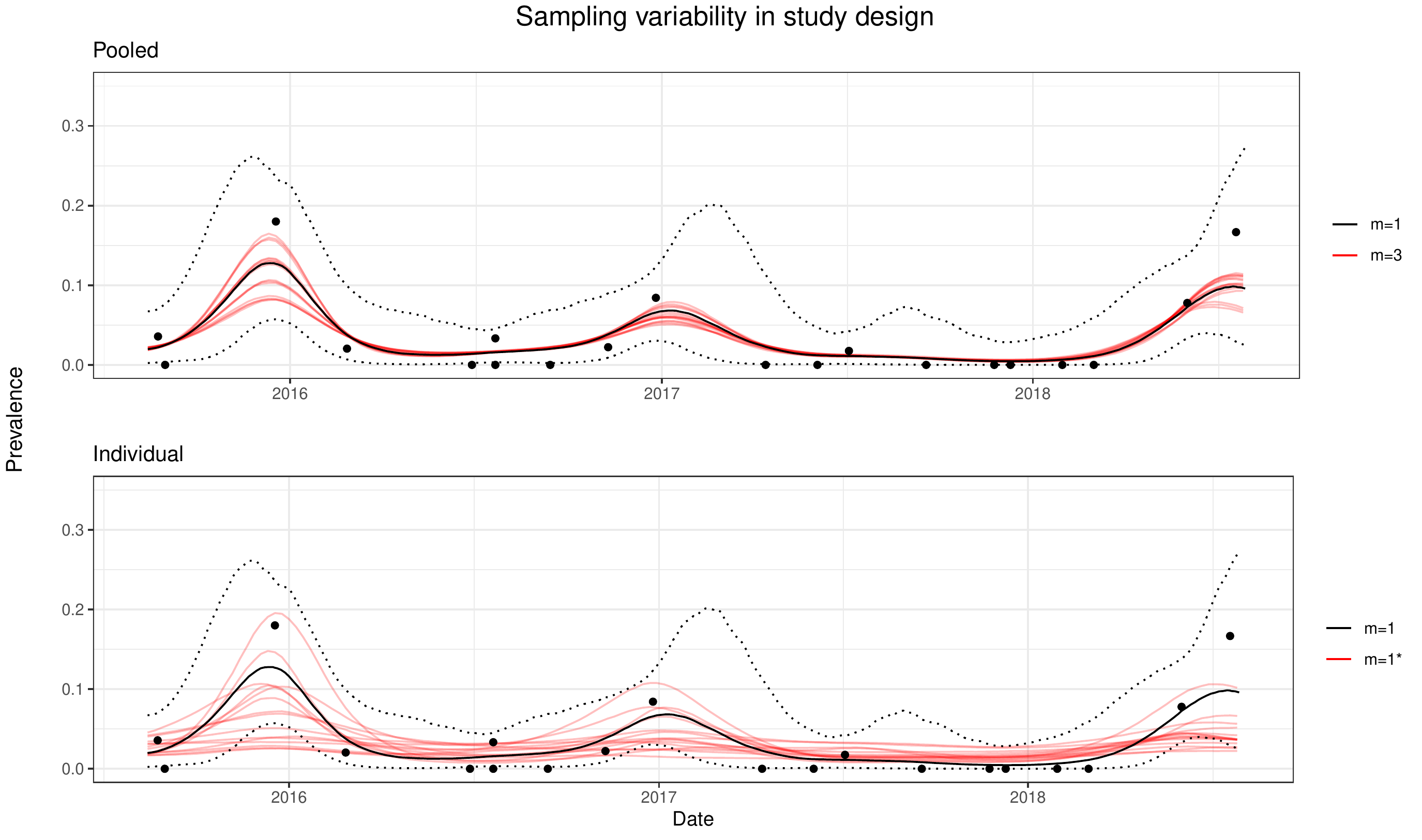}}
\caption{Estimated median curves for 25 resampled pool configurations and 25 subsampled individual configurations are displayed over the estimated median from all available date and corresponding 95\% credible region (identical to Figure \ref{fig2}, $m=1$). Pooled estimates are relatively stable under reconfiguration, but estimates from subsampled individual data are quite sensitive to the subsampling procedure. \label{fig4}}
\end{figure}

Reported results from the Congo Basin bat and Notre Dame COVID-19 surveillance studies are based on sampled pool configurations and subsampled individual results. Other configurations and subsamples are possible, and the sampling variation introduced by this design is not formally represented in the reported results. The Notre Dame study results entail less sampling variability because the sample size is two orders of magnitude larger. We evaluate the impact of the design on the Congo Basin bat example through resampling. 25 pooled ($m=3$) and 25 susbsampled individual data sets are resampled from the complete bat data set and curves are fit to each. Median curves, along with the original $m=1$ curve and its 95\% credible interval, are reported in Figure \ref{fig4}; this figure illustrates variation induced by the pooled sampling assignment. 

In practice, pooling is not achieved by subsampling individual results. Instead, pooled testing either incorporates information from more individuals than would otherwise be tested or tests all individuals at a lower cost. Figure \ref{fig4} is included for completeness and suggests that the pooled results displayed in Figure \ref{fig2} are typical under the pool sampling design. Individual results at the same budget, however, are far more variable. Consider a scenario where $k\times m$ individuals could be sampled, but only $k$ tests are budgeted. Pooled testing permits the incorporation of information from all $k\times m$ individuals and subsequent analyses provide stable approximations to estimates generated from universal individual testing data. 

\subsection{Conclusion}

The simulation studies have established that pooled testing data together with latent GP regression can be used to estimate prevalence over time. The case studies further established that pooled estimates closely match the far more costly alternatives ($m=1$) and that pooled estimates are more precise than budgeted alternative ($m=1^*$). The realized pool assignments and resulting estimates are typical under the pool simulation scheme.

Smoothed and interpolated prevalence estimates are more representative of prevalence dynamics than sets of pointwise estimates. Obtaining such estimates requires more intensive sampling and careful inference. The inferential efficiency of pooled testing, which had previously been shown only in static settings, is seen to apply to dynamic modeling as proposed in this article also. 

\section{Discussion}

Population prevalence as a function of time is estimable from pooled samples. Compared to individual testing, pooling serves to greatly reduce testing costs without qualitatively affecting estimates or uncertainty; at a fixed budget, pooling generates more precise estimates. Applied to surveillance of reservoir host populations, the proposed methodology enables efficient, precise estimation of pathogen prevalence dynamics. 

Throughout, we assumed that unlimited, universal testing best approximates true prevalence, among the strategies considered. A two-stage design is an interesting intermediary between pooled testing and individual testing. It would likely be more costly than pooling and less accurate than individual testing but may achieve a more favorable cost/accuracy trade-off than either. We leave this for future work.

Other natural extensions to this work exist, some of which immediately resolve current limitations. These extensions belong to three thematic categories: incorporating additional information, using estimates to inform sampling, and elaborating on the chosen inference tools.

Known values of sensitivity and specificity may additionally be incorporated at a commensurate expense to precision. A limitation of pooling is the potential for dilution-induced changes in sensitivity and specificity as a function of pool size. Resolving this entails a trade-off between cost and precision. Similarly, modeling covariance jointly in space and time would be an exciting extension applicable to sufficiently rich data sets. 

In online learning settings, where data are analyzed as they become available, various forms of adaptive sampling may be used to design optimal pool sizes at subsequent times. It may also be advantageous to tune sampling frequency in real-time to ensure that dynamics of interest are identified. These adaptive sampling strategies would ensure testing resources are efficiently and effectively expended.

Lastly, the nonparametric tools described in this article were used with relatively basic specifications to establish the general applicability. The model could be strengthened by prior-encoded domain knowledge, alternative covariance functions, and covariate-informed mean structure, depending on the application. Furthermore, as the model is expanded to handle more information, efficient or approximate computation will become increasingly relevant. We look forward to these extensions as exciting future work. 

\section*{Author Contributions}
AH and BS developed the proposed methodology. BS formulated the model, performed analyses, and led writing of the manuscript. AP and RP developed the application setting. AH supervised the project. All authors contributed to drafts and gave final approval for publication.

\section*{Data Availability}
Data and code used in this work are available at \url{https://github.com/braden-scherting/temporal_prevalence}

\bibliographystyle{unsrt}  
\bibliography{references}

\end{document}